\documentstyle{article}
\begin{document}
\title{Distribution functions for hard thermal particles in QCD}
\author{J Frenkel \footnote{This work was supported in part by CNPq and
FAPESP, Brazil}\\{\em Instituto de F\'{i}sica, Universidade de S\~{a}o Paulo,}\\
{\em S\~{a}o Paulo, Brazil} \\ J C Taylor\\{\em DAMTP, Centre for Mathematical Sciences,
Cambridge University},\\{\em  Cambridge, UK}}

\maketitle

\centerline{\em Abstract}

We find a closed-form for the distribution function (defined in terms of
a Wigner operator) for hot coloured particles in a background gluon field,
in the hard thermal loop approximation. We verify that the current is
the same as that derived from the known effective action.

\section{Introduction}

Thermal QCD is beset with IR divergence difficulties. A necessary
first step is to sum the ``hard thermal loops" --- loop diagrams with
external momenta small compared to the temperature $T$ \cite{hardloops}.
These hard thermal loops have been summed up 
 to give an effective action
\cite{effectiveaction}. In the hard thermal approximation, the hot
particles (quarks or gluons) behave semi-classically; and this
suggests the use of a transport equation. Such equations have
been written down by treating the hot quanta as particles \cite{transport},
and also derived by defining the distribution function as the thermal
expectation value of a Wigner operator  \cite{wigner}.

In the second approach, it has been usual to derive the transport
equation from the Wigner operator, and then solve that equation in
the hard thermal approximation. The purpose of this paper is to
omit the intermediate step (the transport equation), and to find a closed
form for the distribution function (that is, the expectation value of the
Wigner operator) in the hard thermal approximation. This closed
form expression ought to encode the same information as the effective
action. We verify this by showing that the current derived from
the distribution function is identical to that coming from the functional
derivative of the effective action.

In order to illustrate our method, we first treat hot scalar ``quarks"
in the background colour field. Then we treat spinor quarks.
Finally, we discuss the case where the hot particles are gluons, using the
background field formalism.

\section{Hot scalar particles}

\def\L{ \Lambda}
\def\p{\partial}
\def\f{\phi}
\def\m{\mu}
\def\n{\nu}
\def\l{\langle}
\def\r{\rangle}
\def\D{{\cal D}}
\def\q{\psi}
\def\g{\gamma}

In order to illustrate our approach as simply as possible, we first
treat the unphysical example of
 a hot scalar field $\f_a$ in the fundamental
representation of $SU(n)$, $a=1,...,n,$ in a slowly varying
background field $A_{\m,ab}$, where we take the $n\times n$ matrix $A$
to be antihermitean and traceless and to include the coupling $g$ in its definition. We work in Lorentz space, with metric $(+,-,-,-)$.

We define a distribution function to be the thermal expectation
value $\l...\r$ (with any zero-temperature part subtracted) 
of a Wigner operator. We use the symbol $W$ to denote this expectation
value.
Thus we define (we denote hermitean conjugation by ${}^*$)
\begin{displaymath}
 W_{ab}(A;x,k)=
\end{displaymath}
\begin{equation}{1\over (2\pi)^4}\int d^4y\exp(ik.y) U_{ac}(A;x,x_+)
\l\f_c(x_+;A)\f^*_d(x_-;A)\r [U^*(A;x,x_-)]_{db},
\end{equation}
where
\begin{equation}
x_+=x+{1\over 2}y,~~~x_-=x-{1\over 2}y,
\end{equation}
and
\begin{equation}
 U(A;x,x_\pm)=P\exp[\int_0^{1/2} du y.A(x\pm uy)]
\end{equation}
and $P$ denotes path-ordering of the matrices $A$.
We have put $A$ in the argument of $\f$ in (1) to emphasize that
$\f$ satisfies the field equation in the background field $A$.
It follows from (1) that $W$ is hermitean.

Under a gauge transformation $\L$,
\begin{equation}
 \phi(x) \rightarrow \L(x)^{-1}\phi(x),
\end{equation}
\begin{equation}
 A_{\m}(x) \rightarrow A^{\L}_{\m}\equiv \L^{-1}A_{\m}\L+\L^{-1}\p_{\mu}\L.
\end{equation} 
Then from (5)
\begin{equation}
 U(x,x_+) \rightarrow \L^{-1}(x)U(x,x_+)\L(x_+) ;
\end{equation}
so that, from (1),
\begin{equation}
 W(x) \rightarrow \L^{-1}(x)W(x)\L(x). 
\end{equation}

To calculate $ W$ in (1), we need solutions of
the field equation for $\f$, in the background field $A$.
This equation is
\begin{equation}
 [\p_{\m}+A_{\m}]^2\f=-m^2\f
\end{equation}
(we give the scalar ``quarks'' a 
mass $m$, since this costs very little extra complication).
We seek an approximate solution of the form 

\[
\f_a=(2\pi)^{-3/2}\int d^4k \delta^+(k^2-m^2)
\]
\begin{equation}
\times[V_{ab}(A;x,k)\exp(-ik.x)a_b(k)+V_{ab}(A;x,-k)\exp(ik.x)b^*_b(k)] 
\end{equation}
where $a$ is an annihilation operator and $b^*$ is a creation operator
(normalized to $[a_a({\bf k}),a^*_b({\bf k'})]=2|{\bf k}|\delta^3({\bf k}-{\bf k'})\delta_{ab}$).
 
 We assume that $k$ is large compared to $A$, and to the rates
 of variation of $V$ and $A$. Then (9) satisfies (8) 
provided that
\begin{equation}
 [2ik.\p+2ik.A-(\p+A)^2]V=0.
\end{equation} 
Write
\begin{equation}
V=V_0+V_1+..., 
\end{equation}
where (defining the covariant derivative $D$)
\begin{equation}
 (k.\p+k.A)V_0\equiv k.DV_0=0,
\end{equation}
\begin{equation}
2ik.D V_1\approx (\p+A)^2V_0\equiv D^2V_0.
\end{equation}
A unitary solution of (12) is
\begin{equation}
 V_0(A;k,x)=P\exp \left[-\int_{-\infty}^0du k.A(x+uk)\right]=1-(k.\p)^{-1}(k.A)+....\end{equation}
We assume that $k.\p$ has a unique inverse when acting on functionals of
$A$ (this is discussed in equation (1.7) of \cite{uniqueinverse});
so that we also have
\begin{equation}
V_0(A;k,x)=\bar{P}\exp\left[\int_0^{\infty}du k.A(x+uk)\right]=[V_0(A;-k,x)]^{-1}. 
\end{equation}
From (13)
\begin{equation}
V_1\approx (2ik.D)^{-1}D^2V_0,
\end{equation}
\begin{equation}
V_1^*\approx - V_0^*D^{*2}(2ik.D^*)^{-1},
\end{equation}
where we make the convention that differential operators like $D^*$
 act to the left.

Using the solutions  (9), and taking the thermal expectation value
of the creation an annihilation operators,
we get from (1) 
\clearpage
\[
 W_{ab}(x,k) ={1\over (2\pi)^4}\int d^4y e^{ik.y}\int d^4 k'\delta^+(k'^2-m^2)N(|k'_0|) \]
\begin{equation}
\left\{e^{-ik'.y}[U(x,x_+)V(x_+,k')
V^*(x_-,k')U^*(x,x_-)]_{ab}~ 
+~(k' \rightarrow -k')\right\},
\end{equation}
where $N$ is the Bose function, and the colours of the quarks in the
heat bath have been summed over.

Now let $B_{\mu}(A;,x,k)$ be
 defined to
be the covariant field which coincides with $A$ in the gauge
where $k.A=0$; so by definition
\begin{equation}
 k.B(A;x,k)=0. 
\end{equation}
This means that there is a gauge transformation, $V_0(A;x,k)$
such that
\begin{equation}
 V_0^{-1}B_{\m}V_0={A^{(V_0)}}_{\m}\equiv V_0^{-1}A_{\m}V_0+V_0^{-1}\p_{\m}V_0
\equiv V_0^{-1}D_{\m}V_0.
\end{equation}
From (19), this requires that $V_0$ should satisfy
(12), which justifies the use of the notation $V_0$.
$B$ satisfies the equation (using $\D$ defined in (26) below)
\begin{equation}
 k.\D B_{\m} \equiv k. \p B_{\m}+k.AB_{\m}-B_{\m}k.A=k_{\n}F_{\n\m}\equiv k.\p A_{\m}-\p_{\m}k.A
+k.AA_{\m}-A_{\m}k.A.
\end{equation}
This is right because it is a covariant equation, and in the gauge
where $k.A=0$ it reduces to $k.\p(B-A)=0$.

In the hard thermal approximation,  $k$ is large compared 
compared to other dimensional quantities. We assume that this means that
$y$ in (1) is small compared to $x$.

We will now separate three contributions to $W$. Type (i) is
leading order, where we neglect $y$ and $V_1$. This term is independent
of $A$, and is simply
\begin{equation}
  W_{(i)ab}={1\over (2\pi)^3} N(|k_0|)\delta(k^2)\delta_{ab}
\end{equation}
as we would expect.

In type (ii), we neglect $ V_1$ in (11) and work to first order in
  $y$ in $U$ and $V_0$. Type (iii) terms
contain $V_1$, and for these we can neglect $y$ in $U$ and $V_0$.

For the  type (ii) term, then, we replace $V$ by $V_0$ in (18), and expand to
first order in $y$, to get
\begin{equation}
 W_{(ii)} ={1\over (2\pi)^4}\int d^4y e^{ik.y}\int d^4k' \delta^+(k'^2-m^2)
 N(|k'_0|)[e^{-ik'.y}R(x,k',y)+
(k' \rightarrow -k')]
\end{equation}
where
\[
R=\left[ y.A(x)+\left((y/2).{\p \over \p x}V_0 \right)V_0^*-V_0\left((y/2).
{\p \over \p x}V_0^* \right)\right
 ]
\]
\begin{equation}
= y.B(x,k';A),
\end{equation}
where we have used (20) together with $V_0^*V_0=1$.

So (23) gives
\begin{equation}
W_{(ii)}={-i\over (2\pi)^3} {\p \over \p k_{\m}}[\delta(k^2)N(|k_0|)B_{\m}(x,k;A)] ,
\end{equation}
where we have used  that $B(-k)=B(k)$ from (21).

In what follows,
we will now find it useful to define a ``2-sided" covariant derivative ${\D}$,
acting on matrices, by
\begin{equation}
 \D_{\m} X\equiv [D_{\m},X]={\p \over \p x_{\m}}X+[A_{\m},X]. 
\end{equation}
If we let $\D$ act on the product $XY$ of two matrices (with the same
argument $x$), we get from (26)
\begin{equation}
\D_{\m}[X(x)Y(x)]=[D_{\m}X]Y+X[YD^*_{\m}],
\end{equation}
where we make the convention that the conjugate operator $D^*$ acts
to the left (as a differential operator), so that
\begin{equation}
 Y\D*_{\m}\equiv [D_{\m}Y^*]^*.
\end{equation}
Note that, in spite of the notation, ${\D}$ is not an $n\times n$ matrix differential
operator acting from the left. Rather it is  an $n^2 \times n^2$
matrix differential operator; so that, acting on a matrix function $X$,
\begin{equation}
 [\D_{\m}X]_{ab}= (\D_{\m})_{ab,cd}X_{cd}.
\end{equation}
Note that, from (21),
\begin{equation}
k.\D(B_{\m}-A_{\m})=-\p_{\m}k.A. 
\end{equation}

We shall require the inverse of the operator $k.\D$, which we will write
shortly as $(k.\D)^{-1}$.
This is an $n^2 \times n^2$ non-local operator. More explicitly:
\begin{equation}
k.\D\l x,ab|(k.\D)^{-1}|x',a'b'\r=\delta^4(x-x')\delta_{aa'}\delta_{bb'}
\end{equation}
and, for any matrix $X(x)$,
\begin{equation}
[ (k.\D)^{-1}X]_{ab}(x)=\int d^4x' \l x,ab|(k.\D)^{-1}|x',cd\r X_{cd}(x')
\end{equation}

Now turn to type (iii) terms. In these, we may set $y=0$ in $U$ and $V_0$.
So, from (16), (17) and (18), we require $S$ where
\begin{equation}
2S(k')= [(k'.D)^{-1}D^2V_0]V_0^*-V_0[V_0^*D^{*2}(k'.D^*)^{-1}],
\end{equation}
where as usual $D^*$ is understood to act to the left.
From (27),
\begin{equation}
2k.\D S=(D^2V_0)V_0^*-V_0(V_0^*D^{*2})=2\D_{\mu}Y_{\mu} 
\end{equation}
where
\begin{equation}
 2Y_{\m}=(D_{\mu}V_0)V_0^*-V_0(V_0^*D^*_{\m})=2B_{\m},
\end{equation}
(using (20)). Hence from (23) and using  $S(-k)=S(k)$,
\begin{equation} 
W_{(iii)}={-i\over (2\pi)^3}(k.\D)^{-1}\D .B\delta(k^2)N(k_0).
\end{equation}

Combining the contribution from (36) with the type (ii) contribution (25),
we finally get
\begin{equation} 
 W_{(ii)}+W_{(iii)} ={-i\over (2\pi)^3}\left[\left({\p \over \p k_{\m}}B_{\m}N(|k_0|)\delta(k^2)\right)+(k.\D)^{-1}\D.BN\delta(k^2)\right]
\end{equation}
This is our result for the distribution function in the hard thermal loop
approximation.

The current is
\begin{equation}
J_{\m}(x)=g\int d^4k k_{\m}W(x,k) =g\int d^4k\theta (k_0)k_{\m}[W(x,k)-W(x,-k)]
\end{equation}
being, like $W$, an $n \times n$  covariant matrix.
Inserting (37) and integrating by parts,
 the temperature-dependent part of the current is
\begin{equation}
J_{\m}(x)={i\over (2\pi)^3}int d^4kN(|k_0|)\delta(k^2)[B_{\m}-k_{\m}(k.\D)^{-1}\D.B].
\end{equation}
Note that
\begin{equation}
 \D_{\m}J^{\m}=0,
\end{equation}
and $J$ is hermitean.

We now claim that this current is the functional derivative
\[ ig{\delta \Gamma \over \delta A^{\m}(x)} \]
 of an effective
action (see equation (29) of\cite{effectiveaction})
\begin{equation}
 \Gamma= -{1\over 2(2\pi)^3}\int d^4x\int d^4k N(|k_0|)\delta(k^2)\hbox{tr}[B(x,k;A)]^2. 
\end{equation}
To verify this,
we work out the variation of (37). From (21), this must have the form
\begin{equation}
\delta B_{\m}(x)=\delta A_{\m}(x)+[f_{\m}k.\delta A](x),
\end{equation}
where $f$ is some (nonlocal) functional of $A$. Therefore
\begin{equation}
 \delta [B^2] =2[B_{\m}+k_{\m}B.f]\delta A_{\m}. 
\end{equation}
But since $B^2$ is an invariant, (43) must vanish for an infintesimal gauge
transformation $\omega$ (using (5) and the definition (26))
\begin{equation}
\delta A_{\m}=\D_{\m}\omega;
\end{equation}
so
\begin{equation}
k.\D B.f= -\D.B
\end{equation}
or
\begin{equation}
B.f=-[k.\D]^{-1}\D.B,
\end{equation}
and thus (46) inserted into the variation of (41) yields the same as
(39), as claimed. (Note that $\delta[\hbox{tr}B^2]=-2B_{ab}\delta B_{ab}$.)

\section{Dirac quarks.}

We now generalize from scalar quarks $\f$ to Dirac quarks $\q$.
Then (1) becomes
\begin{equation}
 W={1\over (2\pi)^4}\int d^4y e^{ik.y}U\l \q \bar{\q}\r U^*, 
\end{equation}
where we suppress colour indices and spinor indices ($W$ is a matrix
in each sort of index).
Instead of (9), we use
\begin{equation}
\q_{\alpha a}= \sum_r(2\pi)^{-3/2} \int d^4k\delta^+(k^2-m^2)
[\exp(-ik.x)V_{ab,\alpha}^{r}a_b^{(r)}+....],
\end{equation}
where $r$ labels two spin states, $a,b$ are colour indices, $\alpha$
is a Dirac index. We expand
\begin{equation}
V_{ab,\alpha}^{r}= V_{0ab}u_{\alpha}^{(r)}+ V_{1ab,\alpha}^{(r)},
\end{equation}
where
\begin{equation}
 (\g .k-m)u^{(r)} =0.
\end{equation}

Now act on $\q$ in (48) with
\begin{equation}
-(i\g.D+m)(i\g.D-m)= (\g .D)^2+m^2= D^2+m^2+{1\over 2}\g_{\m}\g_{\n}F^{\m \n} 
\end{equation}
and use that $(i\g.D-m)\psi=0$.
In a similar way to equation (13), we find that (suppressing colour
 indices and Dirac indices)
\begin{equation}
 2ik.D V_{1}^{(r)}=(D^2V_0)+u^{(r)}{1\over 2}(F^{\m \n}V_0)
\g_{\m}\g_{\n}u^{(r)}.
\end{equation}

When we work out $ W  $, there is a sum over the spin states $r$,
and we use
\begin{equation}
\sum_r u^{(r)} \bar{u}^{(r)} = (k.\g)+m. 
\end{equation}

All the contributions, except the contribution from the $F$ term in (52),
will be like (37) except for an extra factor $k.\g+m$.

In the contribution from the $F$ term in (52),
because the $V_0$ and $V_0^*$ are not differentiated, we may use $V_0V_0^*=1$.
Thus  we get (using (21))
\begin{equation}
2C\equiv [\g_{\lambda}\g_{\nu}(k.\g+m)-(k.\g+m)\g_{\lambda}\g_{ \nu}]
F^{\lambda \nu}
=-2F^{\mu \nu}k_{\m}\g_{\n}=-2(k.\D)B.\g.
\end{equation}

So finally, instead of (34), we have
\clearpage
\[
 W ={-i\over (2\pi)^3}{\p \over \p k_{\m}}[(B_{\m}N(|k_0|)\delta(k^2-m^2)(k.\g+m)]
\]
\[-{i\over (2\pi)^3}[(k.\D)^{-1}(\D.B)(k.\g+m)-B.\g]N(|k_0|)\delta(k^2-m^2)
\]
\begin{equation}
={-i\over (2\pi)^3}(k.\g+m)\left[{\p\over\p k_{\m}}\{B_{\m}N(|k_0|)\delta(k^2-m^2)\}+(k.\D)^{-1}\D .B
N(|k_0|)\delta(k^2-m^2)\right].
 \end{equation}

The current is now given by, in place of (38),
\begin{equation}
J_{\m,ab}=g\int d^4k \hbox{tr} [\g_{\mu} W_{ab} ], 
\end{equation}
where the trace is over Dirac indices.
Using
\begin{equation}
 \hbox{tr}[(k.\g+m) \g_{\mu}]=2k_{\mu}, 
\end{equation}
 we  get for $J$ the same as (39) but with a factor 2 in front
(there are 2 spin states).

\section{Hot gluons.}

We now turn to hot quantized gluons in a background classical Yang-Mills field.
Call the background field $A$ as in previous sections 
and the quantized field ${\cal A}$, so that the total gluon field is
$ A+{\cal A}$. Now we let the indices
$a,b,...$ stand for the adjoint representation. It may be convenient
to think of $A_{ab\m}$ as an $(n^2-1)\times (n^2-1)$ antisymmetric matrix.
We also define $a_a$ by
\[ {\cal A}_{ab}=gf_{abc}a_c.  \]

The Lagrangian for $a$ is
\begin{equation}
{\cal L}={\cal L}_{qu}+{\cal L}_{nonqu},
\end{equation}
where
\begin{equation}
{\cal L}_{qu}=-{1\over 4}[D_{ab\m}a_{b\n}-D_{ab\n}a_{b\m}]^2-
a_{a\m}F_{ab}^{\m\n}a_{b\n}
\end{equation}
is the part quadratic in $a$ (which is the only part directly relevant
to 1-loop digrams), and ${\cal L}_{nonqu}$ is the rest.
Both parts of ${\cal L}$ are separately invariant under
\begin{equation}
A \rightarrow \Lambda^{-1}(D\Lambda),~~~ {\cal A}\rightarrow \Lambda^{-1}
{\cal A}\Lambda,
\end{equation}
where as usual $D=\p+A$.
The invariance (60) will guarantee that the $W$ we derive transforms
 covariantly as in (6). On the other hand, the
 complete Lagrangian ${\cal L}$ is also invariant under
\begin{equation}
A \rightarrow \Lambda^{-1}A\Lambda,~~~ {\cal A}\rightarrow \Lambda^{-1}(D+{\cal A})\Lambda.
\end{equation}
This gauge transformation justifies the  fixing of the gauge in (58),
for instance by adding the gauge-fixing term
\begin{equation}
- {1\over 2}( D.a)^2. 
\end{equation}

For the Wigner function, there are two possibilities (of the same structure as (1)):
\begin{equation}
w_{\m\n}=-{1\over 2 (2\pi)^4}\int d^4y \exp (ik.y)(Ua_{\m})(x_+)(Ua_{\n})(x_-) 
\end{equation}
and
\begin{equation}
W_{\m\n,\lambda\rho}={1\over 4(2\pi)^4}\lim_{m\rightarrow 0}\left [{1\over m^2}\int d^4y \exp(ik.y)(Uf_{\m\n})(x_+)(Uf_{\lambda\rho})(
x_-)\right ],
\end{equation}
were we have suppressed the group indices, and $f$ is defined as
\begin{equation}
 f_{\m\n}={\cal D}_{\m}a_{\n}-{\cal D}_{\n}a_{\m}.
\end{equation}
(A Wigner function like (63) was intoroduced in \cite{josif}.)
In (64), $m$ is some gluon mass, which is to tend to zero.

Both (63) and (64) are invariant under (60), but
(63) has the disadvantage that it breaks invariance under(61)
; so it
may depend on the gauge-fixing of $a$. For (64), it is not clear how to
carry out the limiting process consistently:
it is well known that there is no smooth limit from a massive
to a massless vector field alone (that is, without additional
scalar particles). For this reason, we abandon $W$ in (64) in this paper.

In calculating the (expectation value of) the Wigner function.
we need  the spin sum of the polarization vectors $e$ (analagous to (53)).
 In a ``physical gauge'', we have
\begin {equation}
 \sum_{r=1,2} e^{(r)}_{\m}e^{(r)}_{\n}(k)=-g_{\m\n}+{k_{\m}n_{\n}+k_{\n}n_{\m} \over k.n} \equiv P_{\m\n},
\end{equation}
where $n$ is an arbitrary vector. If we could use $W$ in (64), dependence on $n$
would disappear. But using  $w$ in (63), a
dependence on $n$ may remain.

The complete quadratic action (59) plus (62)
 may be written
\begin{equation}
 -{1\over 2}(D_{\m}a_{\n})^2-a_{\m}F^{\m\n}a_{\n}. 
\end{equation}
The equation of motion is
\begin{equation}
 D^2a^{\m}-2F^{\m\n}a_{\n}=0. 
\end{equation}
Like (49), we seek an approximate solution of (67) of the form
\begin{equation}
V_{ab,\m}^{(r)}=V_{0ab}e^{(r)}_{\m}+ V^{(r)}_{1ab,\m}+....
\end{equation}
Then, using (68) and steps like (12) and (13),
\begin{equation}
i 2k.{\cal D} V^{(r)}_{1,\m}=({\cal D}^2V_0)e^{(r)}_{\m}+2(F_{\m\n}V_0)
e^{(r)\n}.
\end{equation}
The contributions from the first term ($V_0$) in (69) and the first term
in (70) come out very similar to the scalar and Dirac cases, and give
a contribution to $ w_{\m\n}$:
\begin{equation}
{i\over 2(2\pi)^3} \left\{{\p \over \p k_{\lambda}}[B_{\lambda}N\delta(k^2)P_{\m\n}]+(k.\D)^{-1}\D.B
N\delta(k^2)P_{\m\n}\right \}
\end{equation}
where $P_{\m\n}$ was defined in (66).
The contribution from the ``spin term'', the second term in (70) is
\begin{equation}
{i\over 2(2\pi)^3}{1\over n.k}(k.\D)^{-1}[n_{\lambda}F_{\lambda\n}k_{\m}+k_{\lambda}F_{\lambda\n}
n_{\m}]~+(\m \leftrightarrow \n)
\end{equation}

Neither (71) nor (72) can have physical significance in general, because of their
dependence on the arbitrary vector $n$.
However, the trace $w \equiv w^{\m}_{\m}$ is independent of $n$: the trace of (72) is zero and the trace of (71) is similar to (37). We hope that we may
use this trace with some confidence. So we  take the current to be
\begin{equation}
 j_{\m}=g\int d^4k k_{\m} w=g\int d^4k \theta(k_0) k_{\m}[w(k)-w(-k)], 
\end{equation}
and then we get the expected contribution (39)  from (71) (the gluon field
is self-conjugate, but there are 2 polarization states: the trace of (66)
gives $P^{\m}_{\m}=-2$).

In (73), we may us the property
\begin{equation}
w(-k)=w^{cc}(k)
\end{equation}
where $cc$ denotes complex conjugation. Then all reference to negative
values of $k_0$ in $w$ is removed from (73). In order to prove (74), note
that $i$ appears in (63) and (10) only multiplied by $k$ (since $a$ and $A$
are real). Since $w^*=w$, we can alternatively (using (73)) write (74) as
\begin{equation}
j_{\m ab}=g\int d^4k\theta (k_0) k_{\m}[w_{ab}(k)-w_{ba}(k)].
\end{equation}

\section{Conclusion}

The main results of this paper are the three closed expressions for the
distribution functions: (37), (55) and (71) together with (72). In the last
case, we expect only the trace to have physical significance.
The secondary result is the verification that the currents obtained
from these distribution functions coincide with the current coming from
the known hard thermal effective action.

\end{document}